# Boosting output performance of contact-separation mode triboelectric nanogenerators by adopting discontinuity and fringing effect: experiment and modelling studies

Teresa Cheng[1,2], Han Hu[1], Navid Valizadeh[1], Qiong, Liu[1], Florian Bittner[3], Ling Yang[4], Timon Rabczuk[5], Xiaoning Jiang[6], Xiaoying Zhuang[1,7*]

[1] Institute of Photonics, Faculty of Mathematics and Physics, Leibniz University Hannover, Hannover, 30167, Germany

[2] School of Mechanical and Manufacturing Engineering, University of New South Wales, Sydney, New South Wales, 2052, Australia

[3] Institute of Plastics and Circular Economy (IKK), Leibniz University Hannover, Garbsen, 30823, Germany

[4] Geriatric Department, Shanghai Fourth People's Hospital Affiliated to Tongji University, Shanghai, 200434, China

[5] Institute of Structural Mechanics, Bauhaus University, 99423, Weimar, Germany

[6] Department of Mechanical and Aerospace Engineering, North Carolina State University, Raleigh, North Carolina 27695, USA

[7] College of Civil Engineering, Department of Geotechnical Engineering, Tongji University, Shanghai, 20092, China

*Corresponding author: zhuang@iop.uni-hannover.de

## Abstract

Triboelectric nanogenerators (TENGs) are promising self-powering supplies for a diverse range of intelligent sensing and monitoring devices, especially due to their capability of harvesting electric energy from low frequency and small-scale mechanical motions. Inspired by the fact that contact-separation mode TENGs with small contact areas harvest high electrical outputs due to fringing effect, this study employed discontinuity on the dielectric side of contact-separation mode TENGs to promote fringing electric fields for the enhancement of electrical outputs. The results reveal that the TENGs with more discontinuities show higher overall electric performance. Compared to pristine TENGs, the TENGs with cross discontinuities increased the surface charge by 50% and the power density by 114%. However, one should avoid generating discontinuities on tribonegative side of TENGs using metal blade within a positive-ion atmosphere due to the neutralization through electrically conductive metal blade. The computational simulation validated that the TENGs with discontinuities obtained higher electrical outputs, and further investigated the effect of discontinuity gap size and array distance on TENGs performance. This study has provided a promising method for the future design of TENGs using discontinuous structures.

## 1 Introduction

Nanogenerators have recently attracted much attention for energy harvesting due to their ability to convert mechanical energy associated with small-scale physical motions into electric power. Since the first proposal in 2012 [1], triboelectric nanogenerators (TENGs) have become a promising method because of their wide adaptability ranging from walking, running, and clapping of human body, to ocean waves. Compared to another popular nanogenerator - piezoelectric nanogenerators, TENGs outperformed in charging capacitors and harvesting at low frequencies (<4 Hz) [2]. The harvested energy from TENGs can be used to enable various self-powering applications, including wearable devices, wireless sensors, and ultimately implantable electronic devices.

Despite the high output voltage of TENGs, their output current is relatively low, and hence, the output power has potential to improve. For example a fluorinated ethylene propylene (FEP) based TENGs could generate the output voltage of 191 V, but the short-circuit current was reported only 1.4 μA, resulting in the maximum power density of 263.4 mW/m$^2$ [3]. To improve the output power of TENG, various methods have been conducted including materials selection [4], surface microstructures [5], and contact intimacy improvement [6], etc. These methods have boosted the electrical output performance of TENGs significantly, for instance, reaching the power density of 9.9 W/m$^2$ by plasma surface modification of a polytetrafluoroethylene (PTFE) based TENGs [7]. Such power densities have potentially enabled TENGs to become alternative to batteries in low-power devices ranging from 300 nW to 100 µW [8–10], such as pacemakers which only need maximum power of 25 µW (nominal voltage of 3.25V and current consumption of 7.67 µA at 100% pacing) according to Medtronic Azure manual [11].

For contact-separation mode TENGs, multiple researchers have reported that their electrical outputs by unit area (e.g., current density and power density) decrease as the total contact area of TENGs increase [12,13]. It is mainly due to the fringing effect on the edge of TENGs, where the electric field cannot suddenly decrease to zero. As a result, the average electric flux density of TENGs with smaller contact areas is higher than that of larger TENGs. Similar to parallel plate capacitors, the fringing effect of TENGs is not neglectable when the ratio of TENG side length over air gap distance is <10 [14]. However, the influence of fringing effect on TENG output has been rarely investigated quantitively. Several experimental research have measured the electrical outputs of TENGs with various sizes but without further discussion on the underlying mechanism. One computational simulation study has taken into consideration of the fringing effect to array TENGs by optimizing the ratio of TENG size and array distance to achieve the highest electrical outputs [15]. To the best knowledge of the authors, there have been no experimental investigation of the fringing effect on the electrical outputs of TENGs.

Herein, we propose to study the effect of fringing effect on TENGs electrical performance. In this study, fluoriertes ethylen-propylen (FEP) copolymer was used as the dielectric side of TENGs, due to its strongest electron affinity among the various fluorine-containing polymers [16]. We first determined the electrical output of TENGs with various sizes, and then introduced discontinuity on FEP films to create fringing electric field on TENGs. Computational simulation was performed to visualize the influence of fringing effect on electrical output of TENGs with/without discontinuity.

# 2 Materials and methods

## 2.1 Materials

FEP film (RCT®-OHL-1833) with thickness of 50 µm was used here (RCT Reichelt Chemietechnik GmbH + Co, Germany). FEP was used as the triboelectric negative layer, with one side coated with a layer of ca. 100 nm thick copper as electrode using a sputter coater (JEOL JFC-1300, Japan) equipped with a copper target (99.99 %, Micro to Nano, Netherlands). The morphology of FEP surface was observed with a Scanning Electron Microscope (SEM) (JEOL JSM-IT510LA). Aluminum foil with the thickness of 70 µm was used as the triboelectric positive layer, as shown in Fig. 1a.

## 2.2 Specimen

For contact area study, TENG samples with fixed thickness were used and the contact area varied from 3 × 3 $cm^2$, 2.5 × 2.5 $cm^2$, 2 × 2 $cm^2$, 1.5 × 1.5 $cm^2$, 1 × 1 $cm^2$, to 0.5 × 0.5 $cm^2$. For discontinuity study, TENG samples with the contact area of 1.5 × 1.5 $cm^2$ were prepared.

## 2.3 Electrical output measurement

The cyclic contact-separation of TENG with specific frequency, distance, and force was controlled by a linear stage (Linmot DM01-37x120F-HP-R-95), as shown in Fig. 1b. The electrical outputs were collected using an oscilloscope (Rigol DS1074Z). The $I_{sc}$, or short-circuit current, was determined using a low-noise current pre-amplifier (Stanford Research System SR570). We measured the $V_{oc}$, or open-circuit voltage, by connecting an external resistor of 1 GΩ, while we directly assessed the $V_{output}$ using a differential probe (PINTEK DP-50). To gauge the charge Q, we employed an electrometer (Keithley 6517B), and for calculating the maximum power $P_{max}$, we multiplied the output current and voltage with various external resistors. The influence of negative ions on TENG electrical outputs were studied by applying a negative ionic hair dryer (Panasonic EH-NA45) on tribonegative surface of TENG, and ion density was measured using mini air ion tester (Kilter KT-401) [17].

## 2.4 Computational modelling

The analytical modelling of the effect of fringing electric field on TENG output was conducted according to parallel capacitor models. The numerical modelling was carried out using finite element method (Supplementary Materials Note 1).

# 3 Results and Discussion

## 3.1 Surface morphologies

Fig. 1c shows the FEP film before and after Cu coating on the bottom side. The Cu coating thickness was determined using the SEM image (Fig. 1d), yielding an average measurement of approximately 115.5 ± 2.5 nm. Fig. 1d shows that Cu coating was coated evenly on the film surface.

### 3.2 Electrical outputs

First, the condition of the maximum electrical energy was determined. The effect of frequency, distance and force on the electrical outputs can be found in the Supplementary Information Fig. S1. Briefly, higher electrical outputs can be gained with higher frequency, longer distance, and higher force, which can be expected. Under the conditions of 2 cm linear stage distance at the frequency of 3.3 Hz under the applied force of 235 N (Fig. 1f), a 3 × 3 $cm^2$ TENG can generate the maximum peak-to-peak value of $I_{sc}$ of 7.2 μA (or short-circuit current density $J_{sc}$ of 0.8 μA/$cm^2$), the maximum peak-to-peak value of $V_{oc}$ of 364 V, that short-circuit charge transfer $Q_{sc}$ of 7.16 nC, and the maximum peak-to-peak value $P_{max}$ of 148 μW (or power density of 164 mW/$m^2$) (Fig. 1g-k). As a non-modified TENG, these outputs are comparable to the results from previous studies on a 3 × 3 $cm^2$ FEP-based TENG: maximum peak-to-peak value of $I_{sc}$ of 1.4 μA with maximum peak-to-peak value of $V_{oc}$ of 191 V and maximum peak-to-peak value of power density of 263.4 mW/$m^2$ [3].

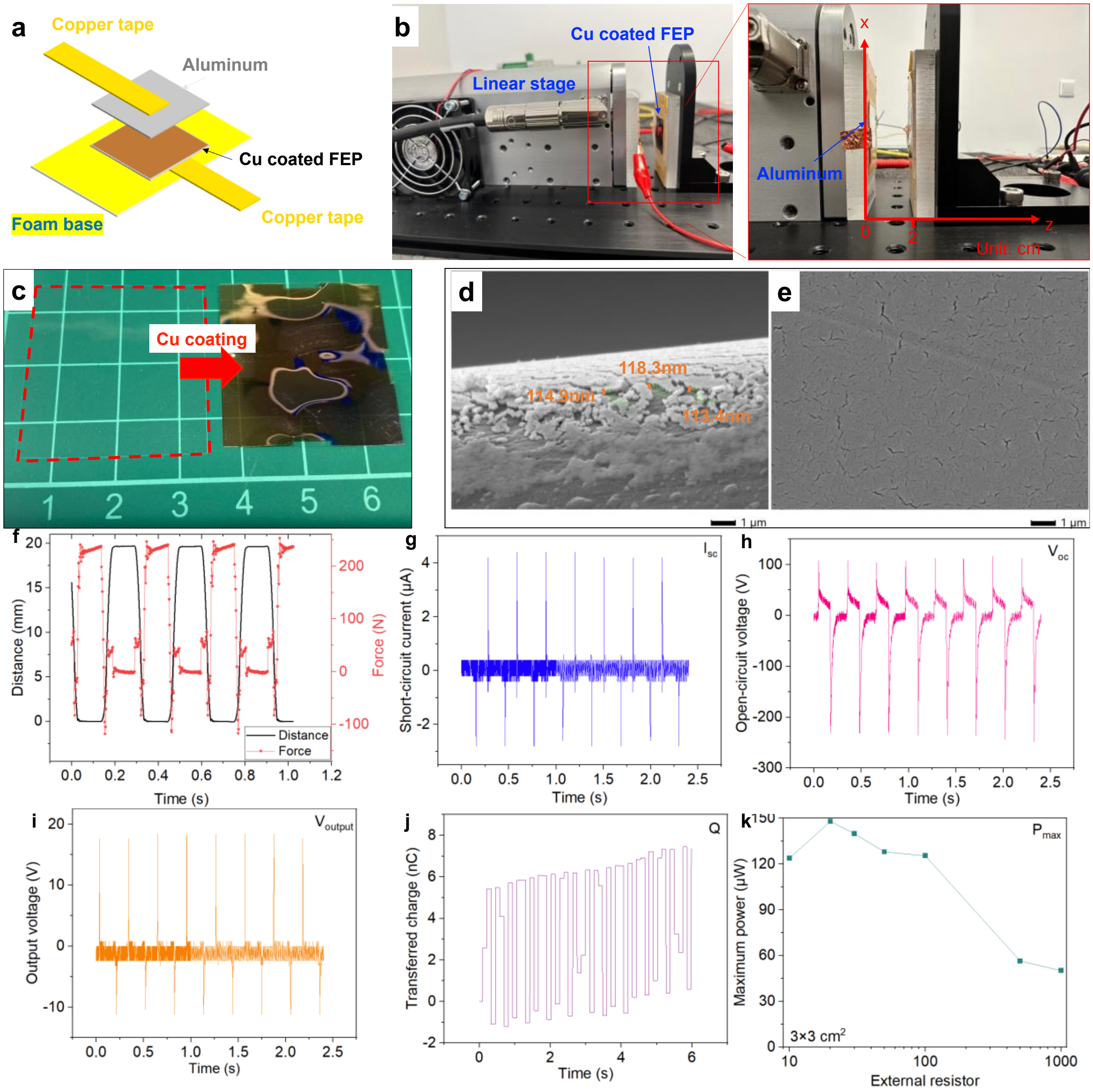

Figure 1. (a) Schematic illustration of a TENG composed of the FEP dielectric side with Cu electrode and the aluminum side. (b) The TENG driven by a linear stage with the contact-separation distance of 2 cm. (c) The pristine FEP film (left) and the FEP film with Cu coated on one side (right). SEM images: (d) the thickness of Cu coating and (e) morphology of Cu coating on FEP surface. (f) The conditions that a 3 × 3 $cm^2$ TENG could harvest the maximum electrical energy - 2 cm linear stage distance at the frequency of 3.3 Hz and the applied force of 235 N: (g) short-circuit current $I_{sc}$, (h) open-circuit voltage $V_{oc}$, (i) output voltage $V_{output}$, (j) transferred charge Q, and (k) maximum power $P_{max}$.

### 3.3 The effect of contact area

The electrical outputs of FEP-based TENGs with various contact areas were measured. The maximum peak-to-peak value of short-circuit current $I_{sc}$ increased from 2.48 μA to 7.2 μA with the increase of contact area size from 1×1 $cm^2$ to 3×3 $cm^2$ (Figs. 2a and b). However, the TENG with a smaller contact area shows the higher short-circuit current density $J_{sc}$: 2.48 μA/$cm^2$ for 1×1 $cm^2$ TENG whilst 0.8 μA/$cm^2$ for 3×3 $cm^2$ TENG. Similarly, although the total transferred charge increased from 2.57 nC to 7.16 nC, the surface charge density σ decreased nonlinearly from 2.57 nC/$cm^2$ to 0.8 nC/$cm^2$ (Fig. 2c and d). The tendencies are aligned with the definition of short-circuit current $I_{sc}$ and current density $J_{sc}$ (Equation 1 and 2), where current and current density are positively dependent on surface charge per area.

$$I_{SC} = \mathrm{A}\frac{d\sigma_u^{SC}}{dt} \quad (1)$$

$$J_{SC} = \frac{d\sigma_u^{SC}}{dt} \quad (2)$$

where A denotes contact area (in $cm^2$), d the air gap distance, $\sigma_u^{SC}$ the short-circuit surface charge density, and t time.

As voltage is also dependent on surface charge density, the maximum peak-to-peak value of open-circuit voltage $V_{oc}$ and the maximum peak-to-peak value of output voltage augmented from 128 V to 364 V (Fig. 2e and f) and from 17.2 V to 29.6 V (Fig. 2g and h) with the contact area rising from 1×1 $cm^2$ to 3×3 $cm^2$, respectively. Consequently, the maximum peak-to-peak value of output power $P_{max}$ increased from 39.2 μW to 148 μW, while the power density decreased from 392 mW/$m^2$ to 164 mW/$m^2$ (Fig. 2i) with an external resistance of 20 MΩ (Fig. 2i inset).

In the meantime, to eliminate the influence of increased applied pressure due to smaller contact area, additional experiment was conducted by maintain a constant area while varying the effective area for electrical outputs. This approach allowed to explore how changes in the contact area affect the conversion of mechanical energy into electrical energy, aligning with the assumption that the mechanical energy depends on the contact area. The results show that the contact area has direct influence on electrical output regardless of the applied force (Supplementary Information Fig. S2). Such negative and nonlinear relation between contact area and output densities (i.e., charge density, current density, and power density) could be due to the fringing effect of electric fields that appear in the edge of electrodes. When the ratio of side length over contact area is higher, the influence of fringing effect is higher. Next, a study on the effect of discontinuity on the output was conducted.

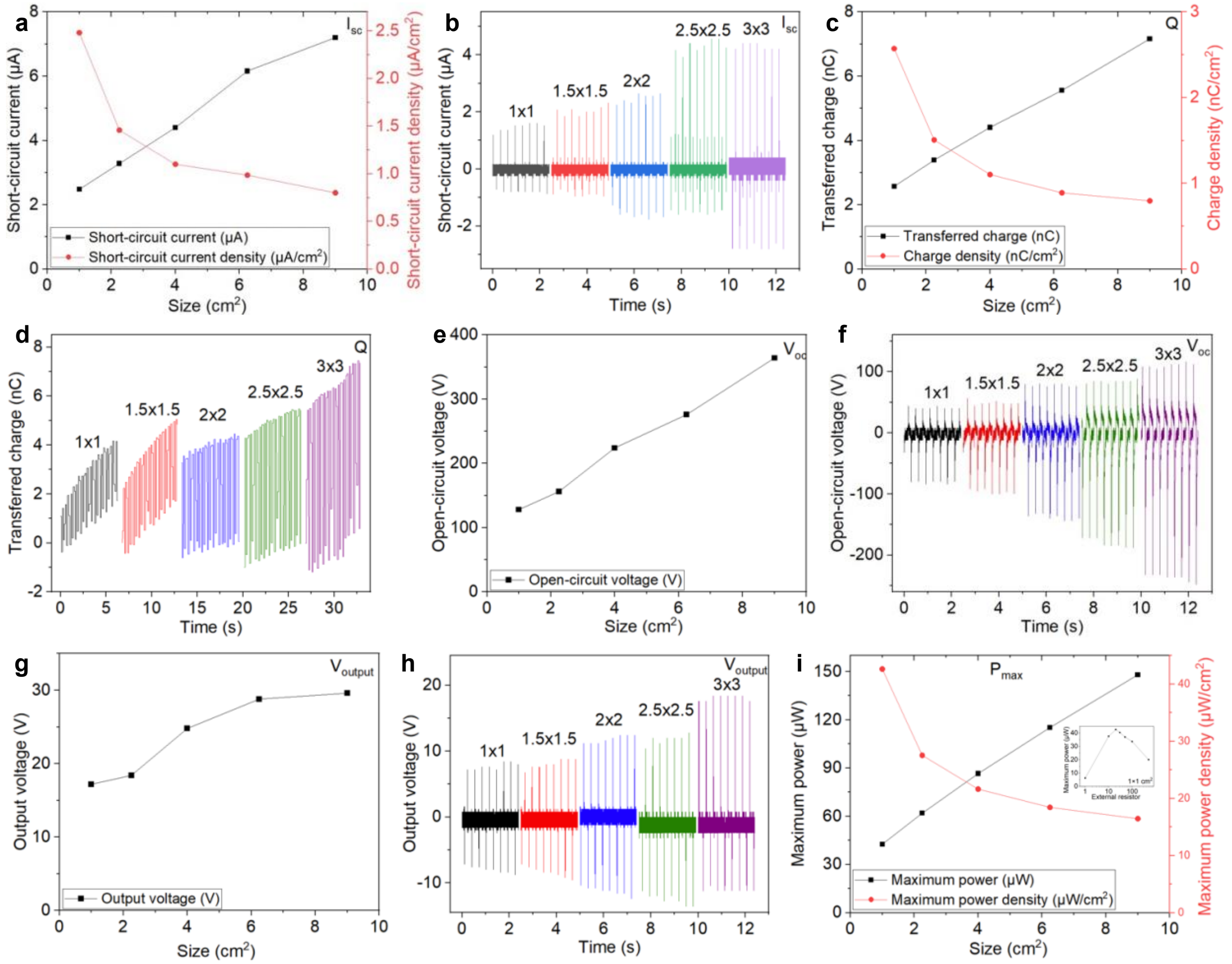

Figure 2. Area-dependent electrical outputs: (a) maximum peak-to-peak value of short-circuit current $I_{sc}$ and current density $J_{sc}$, (b) $I_{sc}$ signals, (c) maximum peak-to-peak value of open-circuit voltage $V_{oc}$, (d) $V_{oc}$ signals, (e) maximum peak-to-peak value of output voltage $V_{output}$, (f) $V_{output}$ signals, (g) maximum peak-to-peak value of transferred charge Q and surface charge density, (h) Q signals, (i) maximum peak-to-peak value of output power $P_{max}$ and power density (inset: the plot of peak power density of 1×1 cm $^2$ TENGs with varying external resistance).

## 3.4 Effect of discontinuity

### 3.4.1 **Discontinuity generated with ceramic blade**

#### *3.4.1.1 Single discontinuity*

To investigate the discontinuity induced fringing effect on electrical outputs, single discontinuity was applied on the FEP side of TENGs. The electrical outputs before cutting (no discontinuity) and after cutting (single discontinuity) were measured, as shown in Fig. 3a. In the meantime, control measurements were conducted as a comparison baseline by monitoring the charge accumulation on pristine TENG surface during the contact separation. Thus, the difference between single discontinuity and the control sample can be regarded as the augmented output due to fringing effect. Three measurements were conducted to obtain the average value along with their standard deviations, and the results were presented as relative values in comparison to the first measurement. Although the control sample shows increase in all electrical parameters compared with the first measurement due to charge accumulation, the output growth of TENG with single discontinuity was higher than the control (Fig. 3 b-f). For example, $I_{sc}$ of TENG with single discontinuity was 1.25 times as high as the TENG prior to cutting (first measurement), while for the control sample, the values were only 1.12 times. By simply subtracting the control sample values from the discontinuity sample values, we could gain that the contribution of fringing effect to the $I_{sc}$ was 13% enhancement. Similarly, the fringing effect led to the increasement of 21% in $V_{oc}$, 7% in $V_{output}$, 18% in Q, and 48% in $P_{max}$. As the same size of TENGs were tested, the increased proportion of power density is identical to the increasement in $P_{max}$, which was 48% in the case of single discontinuity. The results confirmed that regardless of the charge accumulation, single discontinuity could lead to the improvements of overall electrical outputs.

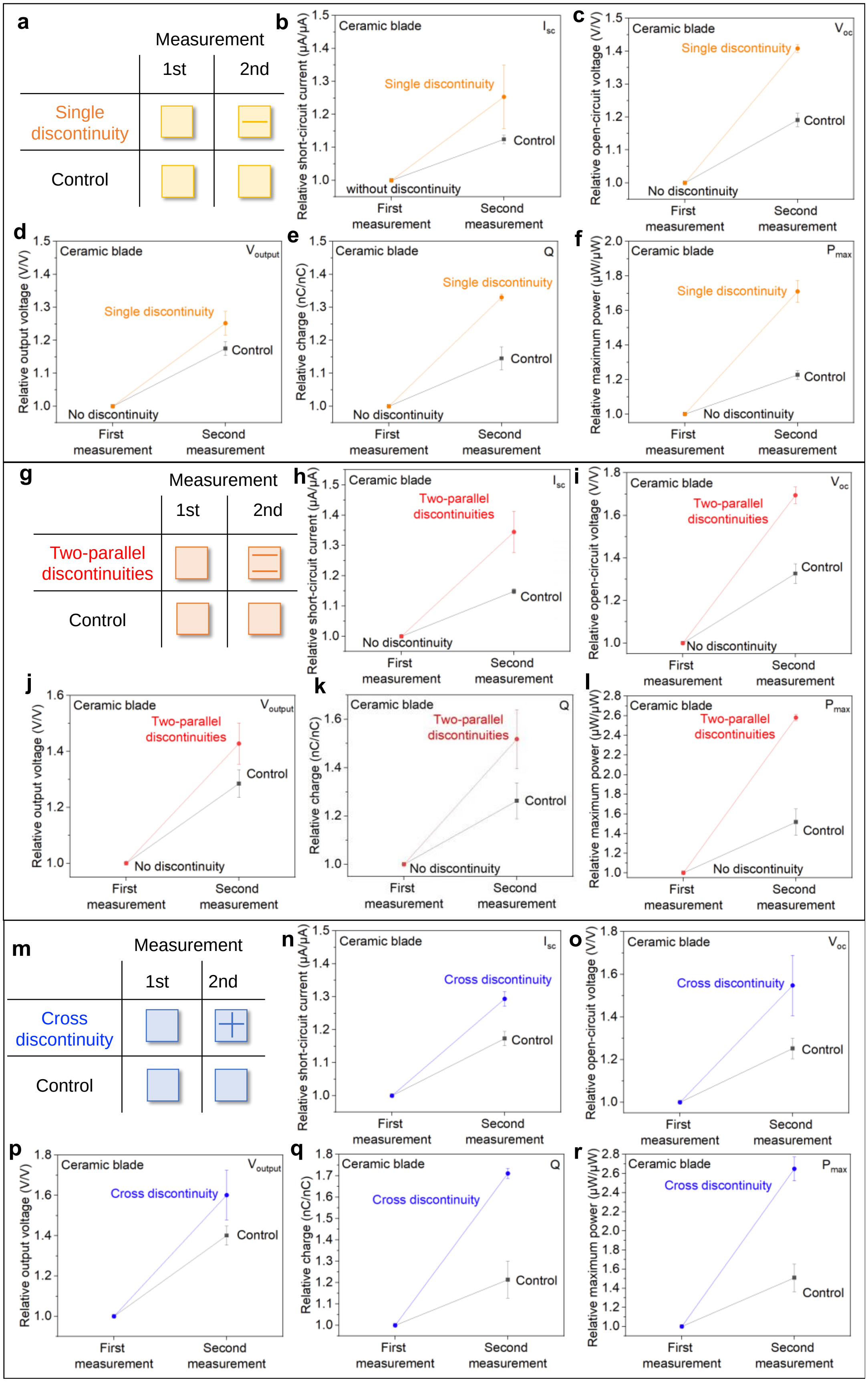

Figure 3. The electrical outputs of TENGs with/without single discontinuity in relative values: (a) illustration of measurement process, (b) short-circuit current $I_{sc}$, (c) open-circuit voltage $V_{oc}$, (d) output voltage $V_{output}$, (e) transferred charge Q, and (f) maximum output power $P_{max}$. The electrical outputs of TENGs with/without two-parallel discontinuities in relative values: (g) illustration of measurement process, (h) short-circuit current $I_{sc}$, (i) open-circuit voltage $V_{oc}$, (j) output voltage $V_{output}$, (k) transferred charge Q, and (l) maximum output power $P_{max}$. The electrical outputs of TENGs with/without cross discontinuity in relative values: (m) illustration of measurement process, (n) short-circuit current $I_{sc}$, (o) open-circuit voltage $V_{oc}$, (p) output voltage $V_{output}$, (q) transferred charge Q, and (r) maximum output power $P_{max}$.

#### *3.4.1.2 Two-parallel discontinuities*

Furthermore, two-parallel discontinuities were applied on the FEP side of TENGs (Fig. 3g). For $I_{sc}$, the presence of two-parallel discontinuities led to an enhancement to 1.34 times of $I_{sc}$ of TENG prior cutting (first measurement), whilst the $I_{sc}$ of control sample increased to only 1.15 times of the first measurement (Fig. 3i). This indicates the fringing effect caused by the two-parallel discontinuities contributed to the enhancement of $I_{sc}$ by 19%. Similarly, the fringing effect led to the increasement of 36% in $V_{oc}$, 14% in $V_{output}$, 25% in Q, and 106% in power density (i.e., $P_{max}$ per $1.5 \times 1.5$ cm$^2$) (Fig. 3i-l). Compared to the outputs of TENG with single discontinuity only, those with two-parallel discontinuities were further enhanced, supporting the assumption that a larger fringing electric field due to more discontinuities could lead to higher electrical outputs.

#### *3.4.1.3 Cross discontinuity*

A more complicated case - cross discontinuity was also investigated (Fig. 3m). By comparing the TENG with cross discontinuity and the control sample in terms of the second measurements, the fringing effect caused by the cross discontinuity led to the increase of 12% in $I_{sc}$, 29% in $V_{oc}$, 20% in $V_{output}$, 50% in Q, and 114% in $P_{max}$ (Fig. 3n-r). These increments did not significantly differ from those with two-parallel discontinuities, suggesting that the fringing effect caused by the two-parallel discontinuities and the cross discontinuity are similar, which will be confirmed by simulation in Section 3.5.

### 3.4.2 **Discontinuity generated with metal blade**

As one of the common cutting tools, a metal blade was also used to create discontinuities on the FEP side of TENGs, and their electrical outputs were measured in comparison to the control sample and the counterpart of ceramic blade (Fig. 4a). It was found that the results were dependent on the environmental conditions (Fig. 4b-f). For example on rainy days, $I_{sc}$ of TENG cut by the metal blade was 1.18 times (single discontinuity, second measurement) and 1.23 times (two-parallel discontinuities, third measurement) as high as the TENG before cutting (no discontinuity, first measurement), and for the control sample, the values were only 1.11 times (second measurement) and 1.15 times (third measurement), indicating the total increase of 7% ~ 8% in $I_{sc}$ due to fringing effect. However on sunny days, the TENG cut by the metal blade show decreasing output to 0.92 times (one cutting) and 0.85 times (two cuttings) as high as the $I_{sc}$ of TENG before cutting, demonstrating the total reduction of 19% for single discontinuity and 30% for two-parallel discontinuities. In the case of cloudy days, the $I_{sc}$ of TENGs cut by metal blade was between the performance on rainy day and sunny day. Such tendency can also be observed in all other parameters: $V_{oc}$, $V_{output}$, Q, and $P_{max}$.

Two assumptions were raised to explain the weather-dependent performance. The first was the difference in relative humidity between a rainy day and a sunny day, which could reach the difference up to 63% [18]. However, it has been proven that 10% relative humidity attributes to only 2.5% enhancement in surface charge density [19], and our experiments using humidifier have also confirmed the insignificant influence of relative humidity on TENG electrical outputs (Fig. S3 in Supplementary Information).

Another assumption was the atmospheric electricity, which means the air is dominant by positively charged ions on sunny days and more likely full of negatively charged ions on rainy days [20–22] (Fig. 4g). To confirm the assumption, negative ions with the density of 1.6 ions/$cm^3$ were generated surrounding the TENG and the electrical output $I_{sc}$ with or without the metal blade were measured. As shown in Fig. 4h, the negative ions alone cannot improve the TENG output, but combined the touching of TENG with the metal blade significantly enhanced the output by 18%. Normally, the FEP (tribonegative) side of TENGs gained negative charges after repeated contact and separation, and there was not a direct exchange of charges between air and FEP surface due to the insulting property of polymers, preventing it from readily exchanging charges with the surrounding air. The conductive metal blade, when in contact with the FEP surface, attracted and carried away the negative charges from the FEP surface (Fig. 4i). The positive ions in the atmosphere then neutralized these charges, leading to a decrease in the negative charge on the FEP surface and overall lower electrical outputs of TENGs. In contrast, on rainy days where the atmosphere is dominant by negative ions, the negative charges on FEP surface would not be neutralized. However, the conductive metal blade would still attract and carry away some negative charges from the FEP surface, resulting in lower electrical outputs than the TENGs cut by non-conductive ceramic blade where more charges remain on the FEP surface.

Such findings have not been proposed previously, possibly because TENGs were tested in closed labs and never exposed to negative ions from outdoor rainy weather. Instead, an environment full of electronic instruments is more likely to have positive atmospheric electricity, thus adding discontinuities on TENGs by a metal blade could not boost the electrical outputs. A non-conductive blade, such as the ceramic blade did not reveal weather-dependent performance (Fig. S4 in Supplementary Information). Therefore, it is recommended to use a nonconductive blade to cut the tribonegative side of TENGs regardless of the atmospheric electricity.

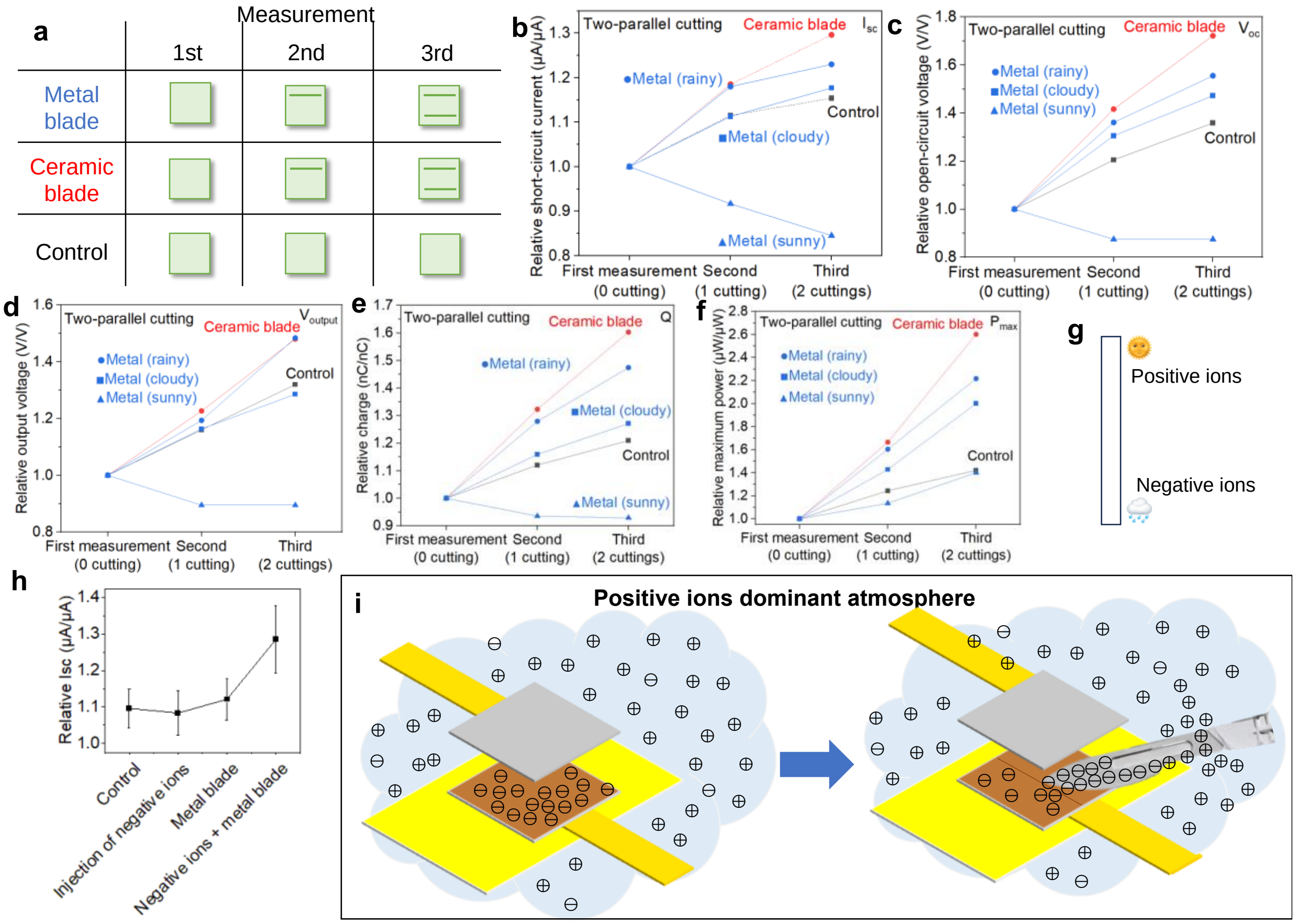

Figure 4. The electrical outputs of TENGs with/without two-parallel cutting by metal/ceramic blade in relative values: (a) illustration of measurement process, (b) short-circuit current $I_{sc}$, (c) open-circuit voltage $V_{oc}$, (d) output voltage $V_{output}$, (e) transferred charge Q, and (f) maximum output power $P_{max}$. (g) The charge of ions tends to be positive on sunny days and negative on rainy days. (h) The electrical output $I_{sc}$ of TENG with the exposure of negative ions and/or metal blade. (i) The schematic illustration of negative charges on FEP surface being attracted and carried away by the conductive metal blade and neutralized with the positive ions on sunny days.

## 3.5 Computational modelling

The computational modelling was employed to study the mechanics behind the effect of discontinuity on boosting TENGs' outputs. By validating the experimental results and simulating a variety of parameters, the modelling can further provide directions to higher output induced by various discontinuities.

### 3.5.1 Constitutive relation

The relationship between electric field $\boldsymbol{E}$ and charge $q$ can be derived by the Coulomb's Law as

$$\boldsymbol{E}(\mathrm{r}) = \frac{q}{4\pi\varepsilon}\frac{(\boldsymbol{r}-\boldsymbol{r}')}{|\boldsymbol{r}-\boldsymbol{r}'|^3} \quad (3)$$

where $\varepsilon$ denotes the constant permittivity of the dielectric material, $\boldsymbol{r}$ is the field point where the electric field is evaluated, and $\boldsymbol{r}'$ is the point source of electric charge. Based on the computational model [23], we have reproduced the simulation of contact-separation mode TENGs in terms of electric potential distribution (Fig. 5a) and further confirmed the fringing electric field on the edges (Fig. 5b-d). When the distance between the TENGs is half of the length/width of TENG, the fringing effect becomes pronounced and cannot be neglected [24]. Multiple fringing electric fields would lead to a non-uniform polarization inside the dielectrics and a larger electric field [15].

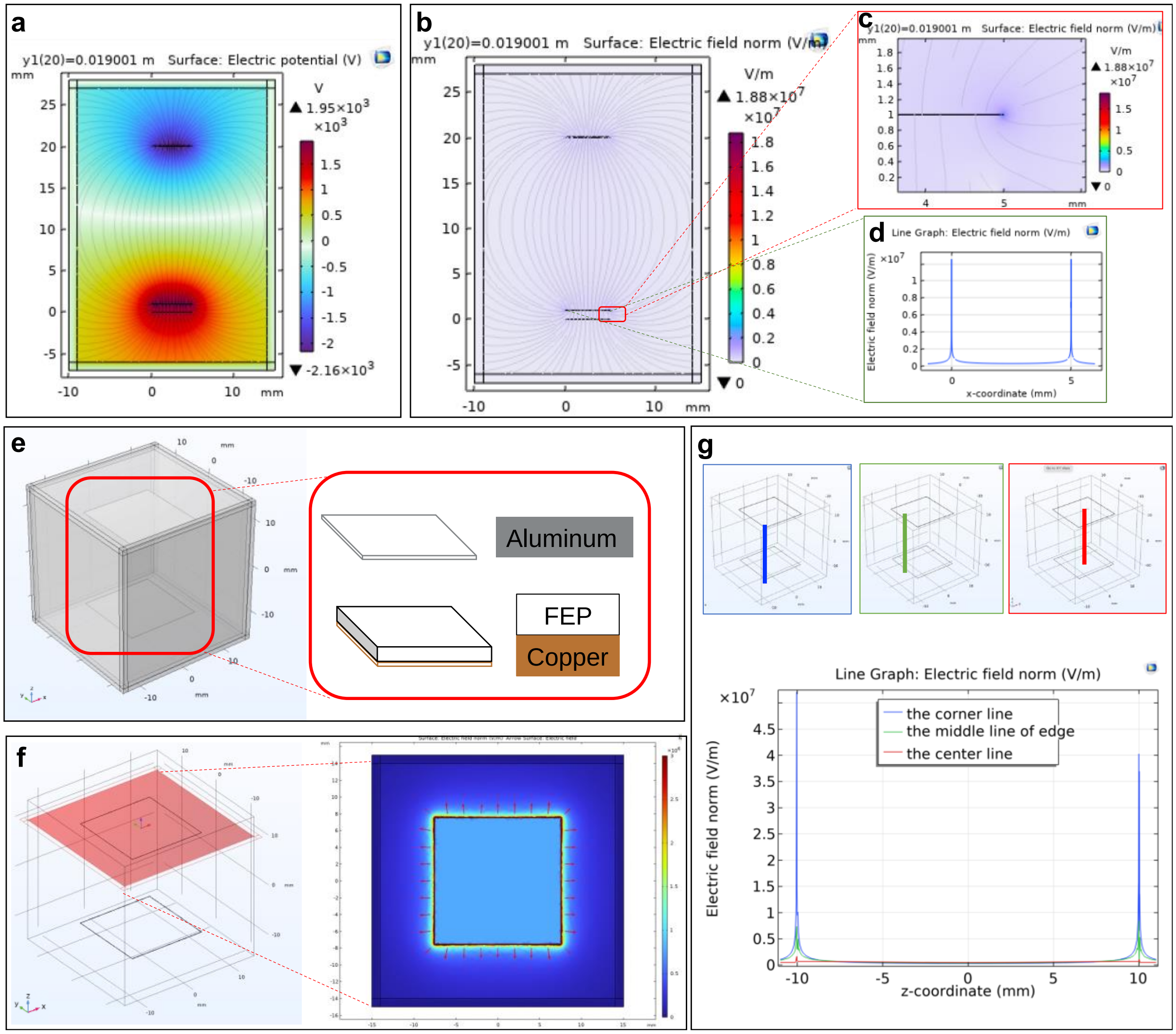

Figure 5. (a) The reproduction of electric potential distribution of contact-separation mode TENGs in reference [23]. (b) Based on the reference [23], the distribution of electric field was further simulated, and the fringing electric field was observed on the edge of TENGs in (c) colors and (d) numbers. (e) The FEM modelling of TENGs consisting of FEP film with copper electrode and aluminum film. (f) The electric field strength on the aluminum surface that shows substantially higher value (in red) on the edge. (g) The electric field along the vertical z direction when z was placed in the center of TENG (in green), in the middle of the edge (in green), and in the corner of TENG (in red).

### 3.5.2 **The FEM model**

The TENG was modelled by using the FEM as shown in Fig. 5b. All the parameters were set according to the experiments in Section 3.4, where the TENG contact area was 1.5 × 1.5 $cm^2$, and the distance between FEP and aluminum parts of TENG was 20 cm. The atmosphere air was a 3 × 3 × 3 $cm^3$ cube, with 0.1 cm thick layer of infinite element domain on the boundaries. The distribution of electric field can be found in Supplementary Information Fig. S5.

The fringing effect was clearly observed on the edges of aluminum side (Fig. 5c), with the electric field strength of ~ 3 × $10^6$ V/m on the edge (red color), in comparison to the strength of ~ 1 × $10^6$ V/m inside the TENG (light blue color). The electric field strength on other sections can be found in Supplementary Information Fig. S6. For a TENG, electric field strength along three vertical lines were compared in Fig. 5d: the value at the intersection of the dielectric surface and the corner line (blue) was 5.2 × $10^7$ V/m (z-coordinate = -1 cm), 0.7 × $10^7$ V/m at the intersection of the dielectric surface and the middle line of edge (green), and 0.1 × $10^7$ V/m at the intersection of the dielectric surface and the center cut line (red). Obviously, the electric field strength along the corner line was significantly higher than other two, suggesting the introduction of discontinuity and fringing effect could lead to enhanced electric fields.

### 3.5.3 **Modelling of discontinuities and energy output**

#### *3.5.3.1 Discontinuity patterns*

Fig. 6 show the electric field strength on the dielectric surface of various TENG scenarios: original entire TENG, a TENG with single discontinuity in the middle of dielectric film, a TENG with two-parallel discontinuities on the dielectric side, and a TENG with cross discontinuity on the dielectric side. The electric field strength was simulated based on the input surface charge. Here, the surface charge Q from experimental measurement was used: 2.03 nC for original TENG, 2.68 nC for single-discontinuity TENG, 3.25 nC for two-parallel- discontinuity TENG, and 3.4 nC for cross-discontinuity TENG. The resulting electric field strength at the corners increased with the introduction of discontinuities. Compared to the original TENG, the single discontinuity led to the rise of electric field strength at the corners from 1.17 × $10^7$ V/m to 1.47 × $10^7$ V/m (Fig. 6a and c), further to 1.73 × $10^7$ V/m for the TENG with two-parallel discontinuities (Fig. 6e), and even to 1.8 × $10^7$ V/m when a cross discontinuity was applied (Fig. 6g). In terms of the middle lines across the dielectric surfaces (Fig.6b, d, f, and h), the peak of electric field strength of original TENG was $6.0 \times 10^6$ V/m, and increasing to $7.4 \times 10^6$ V/m for single-discontinuity TENG, $7.5 \times 10^6$ V/m for two-parallel- discontinuity TENG, and $8.6 \times 10^6$ V/m for cross-discontinuity TENG. Meanwhile, except for the edges, the electric field strength along the middle lines across the dielectric surface (flat parts in Fig.6b, d, f, and h) also increased from $6.7 \times 10^5$ V/m to $8.9 \times 10^5$ V/m by introducing single discontinuity, and reached $1.0 \times 10^6$ V/m for two-parallel- discontinuity TENG and $1.0 \times 10^6$ V/m for cross-discontinuity TENG. Both the peak values and plateau values confirmed that the discontinuities enhanced electrical field strength.

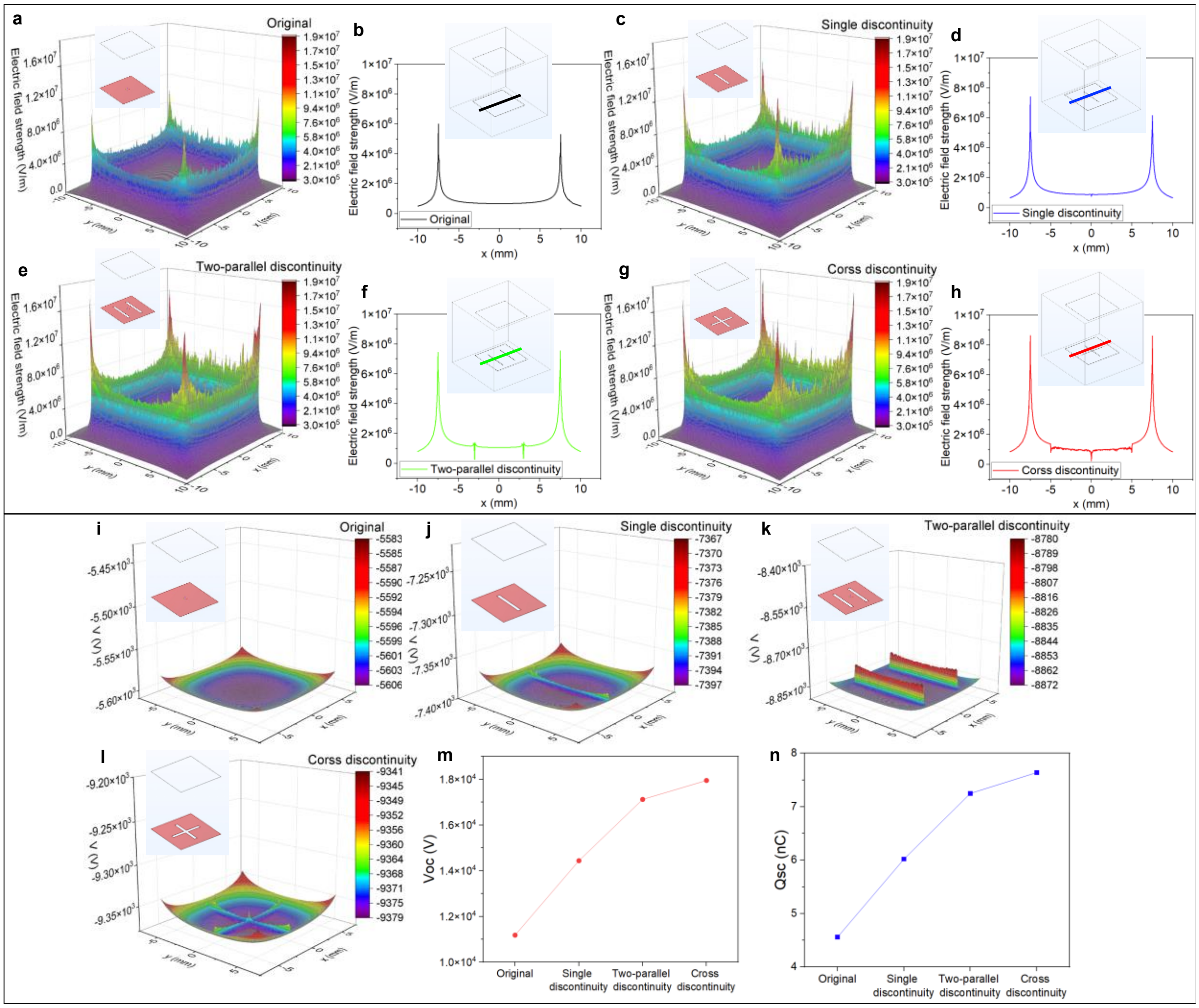

Figure 6. The distribution of electric field strength on the TENG surfaces and along the middle lines across the dielectric surface: (a-b) original TENG, (c-d) TENG with one cutting in the middle of dielectric side, (e-f) TENG with two cutting on the dielectric side, and (g-h) TENG with cross cutting on the dielectric side. The distribution of voltage on the dielectric side of TENG in the open-circuit condition: (i) original TENG, (j) TENG with one cutting in the middle of dielectric

side, (k) TENG with two cutting on the dielectric side, and (l) TENG with cross cutting on the dielectric side. (m) The average $V_{oc}$ of TENGs with various discontinuities. (n) The short-circuit charge transfer $Q_{sc}$ of TENGs with various discontinuities.

Furthermore, the $V_{oc}$ and $I_{sc}$ of TENG with various discontinuities were simulated. The distribution of voltage in the open-circuit condition on all surfaces of TENGs were shown in Fig. 6i-l. The average values of $V_{oc}$ of the TENGs were compared (Fig. 6m), indicating higher $V_{oc}$ was achieved with more discontinuities. Regarding the short-circuit charge transfer $Q_{sc}$ (Fig. 6n), more discontinuities led to higher values, referring to higher $I_{sc}$.

#### *3.5.3.2 Discontinuity gap width*

The effects of discontinuity gap width on TENG electrical outputs were investigated. Taking the TENG with single discontinuity as an example, there was a tendency that a wider gap width led to higher electric potential on the tribonegative surface of TENGs in open circuit (Fig. 7a), indicating lower $V_{oc}$ (potential difference between tribopositive and tribonegative surfaces) generated by the TENGs with larger gap width (Fig. 7e). By comparing the distributions of electric potential on the tribonegative surface (Fig. 7b-d), a smaller potential difference could be seen on the discontinuity gap area of TENG with a gap width of 0.001 mm than that of the TENG with gap width of 0.005 mm and 0.01 mm. Consistently, the short-circuit charge transfer $Q_{sc}$ also suggested a wider discontinuity gap lead to lower electrical outputs (Fig. 7f). Therefore, despite the benefits of introducing discontinuity on TENGs, smaller gap widths should be applied.

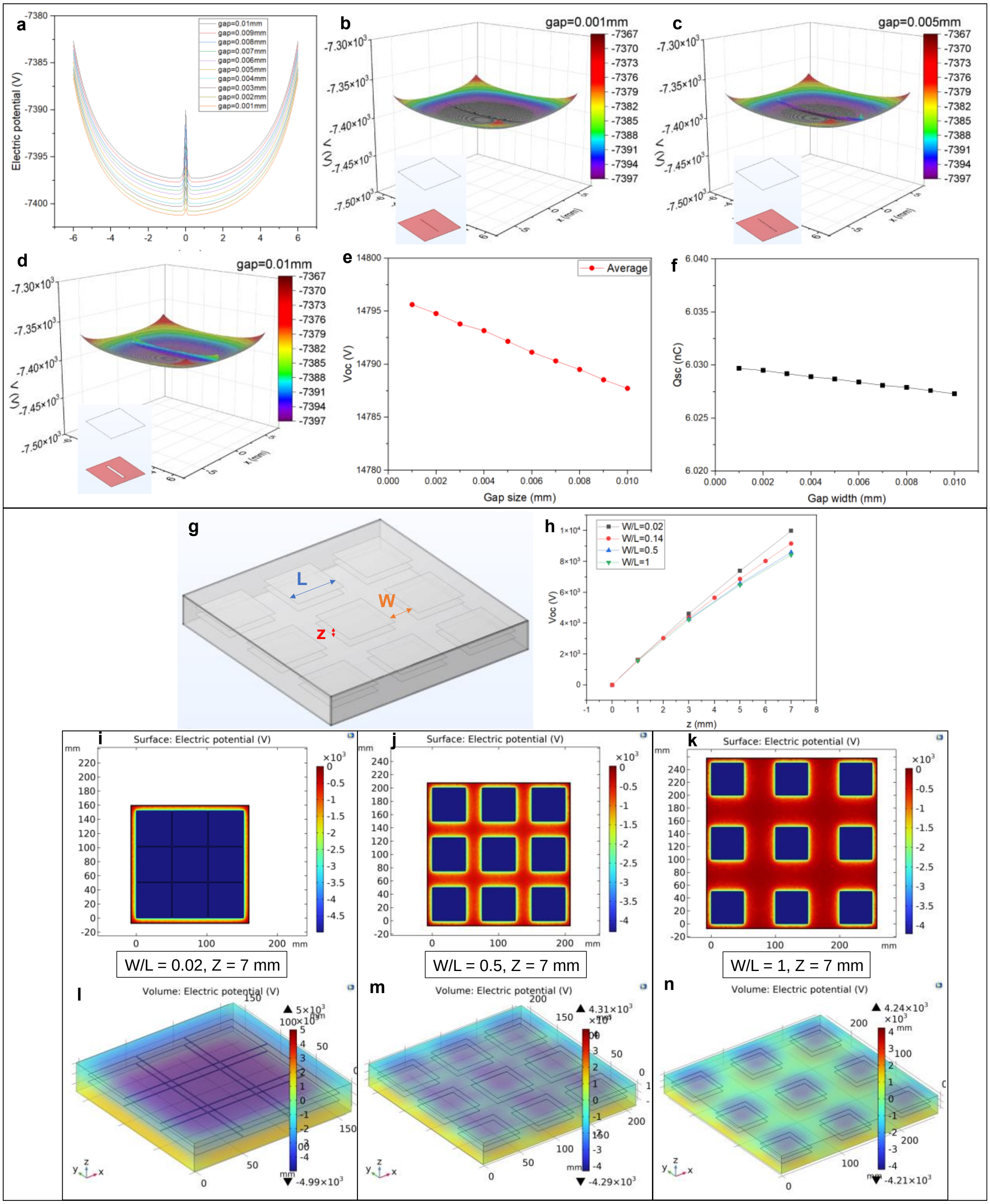

Figure 7. (a) The electric potential on the dielectric surface of TENGs with various single discontinuity gap width. The typical distributions of electric field strength of TENG with single discontinuity: (b) discontinuity gap width of 0.001 mm, (c) discontinuity gap width of 0.005 mm,

and (d) discontinuity gap width of 0.01 mm. (e) The average $V_{oc}$ of TENGs with various discontinuity gap width. (f) The short-circuit charge transfer $Q_{sc}$ of TENGs with various discontinuity gap width. (g) The illustration of TENG array. (h) The $V_{oc}$ as a function of distance between TENGs (z) with various ratios (W/L) of horizontal distance between TENGs (W) and TENG length (L). The electric potential distribution on the dielectric surfaces of TENG arrays with (i) W/L = 0.02, (j) W/L = 0.5, and (k) W/L = 1. The 3D electric potential distribution of TENG arrays with (l) W/L = 0.02, (m) W/L = 0.5, and (n) W/L = 1.

### 3.5.4 **Study of TENG arrays**

An array composed of nine TENGs were investigated in terms of the vertical distance between TENGs (z), and the ratio (W/L) of horizontal distance between TENGs (W) and TENG length (L) (Fig. 7i). As shown in Fig. 7j, larger distance between TENGs (z) led to higher $V_{oc}$, which is consistent with the results reported previously [25]. Additionally, smaller W/L ratios generated higher $V_{oc}$, as supported by the higher electric potential distribution on the tribonegative surface of TENGs with W/L = 0.02 (Fig. 7i and l) than that of W/L = 0.5 (Fig. 7j and m) and W/L = 1 (Fig. 7k and n). Hence, TENG arrays with larger vertical distance but smaller horizontal distance are favourable.

## 3.6 Demonstration

The ability of TENG with or without discontinuity to charge capacitor was measured using the circuit shown in Fig. 8a. First of all, the TENG can harvest enough power to lighten 11 LEDs (Fig. 8b). The time taken to charge a 22 μF were compared, showing the TENG with cross discontinuity has the highest power output, followed by the TENG with single discontinuity alone and the original TENG (Fig. 8c).

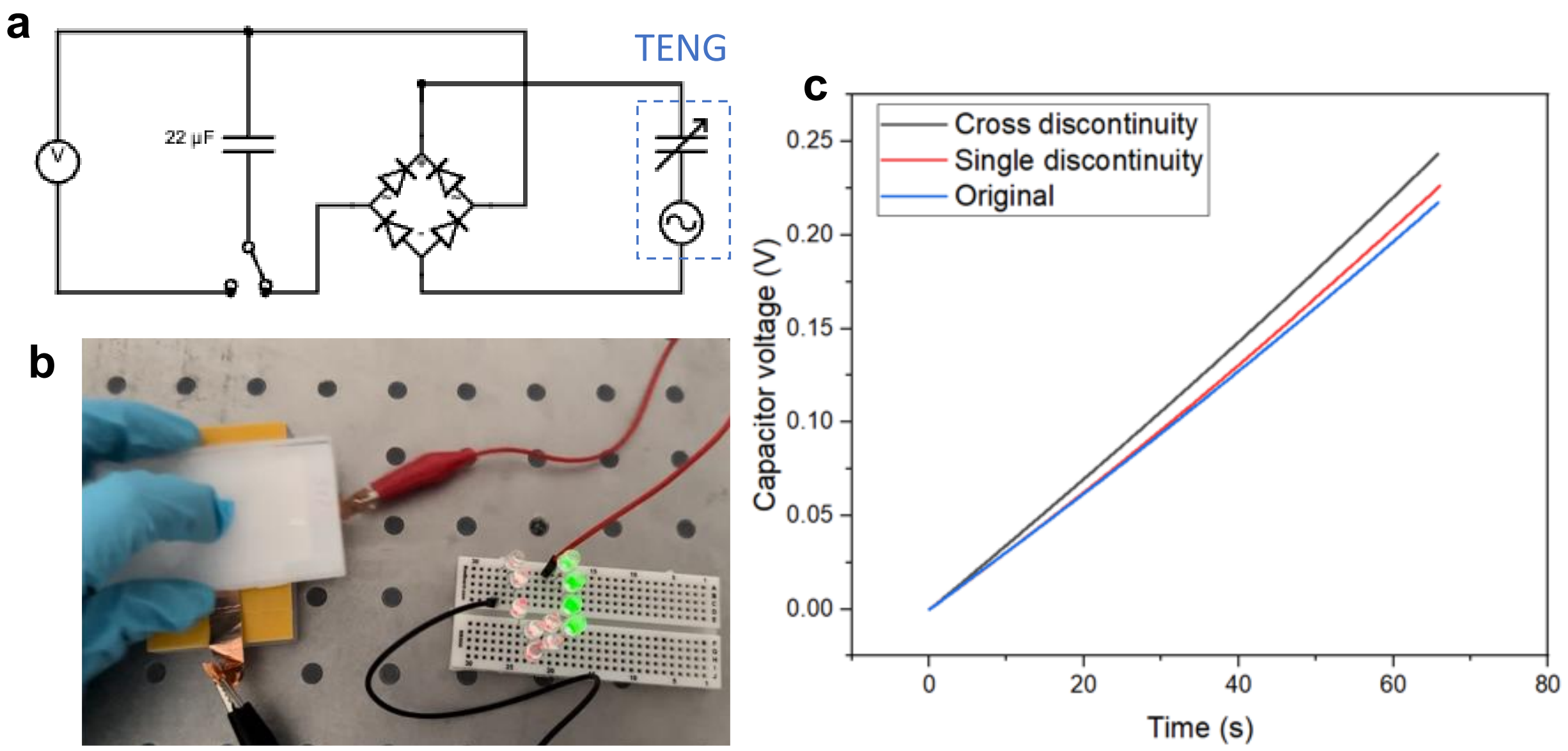

Figure 8. (a) The circuit diagram for charge measurement of TENGs. (b) Demonstration of lightening LEDs using a contact-separation mode TENG. (c) The charging curve of TENGs with or without discontinuity.

## 4 Conclusion

In this study, it was shown that the fringing effect generated by the discontinuity of contact-separation mode TENG surface can boost the electrical outputs. By introducing single, two-parallel, and cross discontinuity, it was found that more discontinuities tend to increase more electrical outputs, reaching maximum value of $P_{max}$ by 114% after introduction a cross discontinuity on the tribonegative side of TENGs. Additionally, atmosphere conditions such as relative humidity and atmospheric electricity also have significant influence on the surface charge density of TENGs, where further research can be conducted to take the advantage of this property for fabrication process. The study employed the computational simulation to validate the enhancement of electrical outputs of TENGs with discontinuities. The results indicated that smaller discontinuity gaps led to higher electrical outputs, making TENG arrays with smaller horizontal distances a favorable choice. However, it remains a subject of ongoing research to fully comprehend why TENGs with discontinuities attract more surface charge compared to pristine TENGs. This intriguing phenomenon warrants further investigation in the future to unlock its potential benefits and applications.

## Acknowledgement

This project has received funding from the European Union's Horizon 2020 research and innovation programme under the Marie Skłodowska-Curie grant agreement No. 101026292. Authors also would like to thank the Cluster of Excellence PhoenixD, ERC Grant (802205), and the Deutsche Forschungsgemeinschaft (DFG, German Research Foundation) – 467965905.

## Reference

[1] F.-R. Fan, Z.-Q. Tian, Z. Lin Wang, Flexible triboelectric generator, Nano Energy. 1 (2012) 328–334. https://doi.org/10.1016/j.nanoen.2012.01.004.

[2] A. Ahmed, I. Hassan, A.S. Helal, V. Sencadas, A. Radhi, C.K. Jeong, M.F. El-Kady, Triboelectric Nanogenerator versus Piezoelectric Generator at Low Frequency (<4 Hz): A Quantitative Comparison, iScience. 23 (2020) 101286. https://doi.org/10.1016/j.isci.2020.101286.

[3] J. Wang, B. Wu, G. Liu, T. Bu, T. Guo, Y. Pang, X. Fu, J. Zhao, F. Xi, C. Zhang, Flexure hinges based triboelectric nanogenerator by 3D printing, Extreme Mechanics Letters. 20 (2018) 38–45. https://doi.org/10.1016/j.eml.2018.01.002.

[4] R. Zhang, H. Olin, Material choices for triboelectric nanogenerators: A critical review, EcoMat. 2 (2020) e12062. https://doi.org/10.1002/eom2.12062.

[5] L. Zhao, Q. Zheng, H. Ouyang, H. Li, L. Yan, B. Shi, Z. Li, A size-unlimited surface microstructure modification method for achieving high performance triboelectric nanogenerator, Nano Energy. 28 (2016) 172–178. https://doi.org/10.1016/j.nanoen.2016.08.024.

[6] J. Wang, C. Wu, Y. Dai, Z. Zhao, A. Wang, T. Zhang, Z.L. Wang, Achieving ultrahigh triboelectric charge density for efficient energy harvesting, Nat Commun. 8 (2017) 88. https://doi.org/10.1038/s41467-017-00131-4.

[7] T. Prada, V. Harnchana, A. Lakhonchai, A. Chingsungnoen, P. Poolcharuansin, N. Chanlek, A. Klamchuen, P. Thongbai, V. Amornkitbamrung, Enhancement of output power density in a modified polytetrafluoroethylene surface using a sequential O2/Ar plasma etching for triboelectric nanogenerator applications, Nano Res. 15 (2022) 272–279. https://doi.org/10.1007/s12274-021-3470-4.

[8] K. Rawy, R. Sharma, H.-J. Yoon, U. Khan, S.-W. Kim, T.T.-H. Kim, A triboelectric nanogenerator energy harvesting system based on load-aware control for input power from 2.4 μW to 15.6 μW, Nano Energy. 74 (2020) 104839. https://doi.org/10.1016/j.nanoen.2020.104839.

[9] M. Magno, N. Jackson, A. Mathewson, L. Benini, E. Popovici, Combination of hybrid energy harvesters with MEMS piezoelectric and nano-Watt radio wake up to extend lifetime of system for wireless sensor nodes, in: 26th International Conference on Architecture of Computing Systems 2013, 2013: pp. 1–6. https://ieeexplore.ieee.org/abstract/document/6468881 (accessed October 13, 2023).

[10] C. Moerke, A. Wolff, H. Ince, J. Ortak, A. Öner, New strategies for energy supply of cardiac implantable devices, Herzschr Elektrophys. 33 (2022) 224–231. https://doi.org/10.1007/s00399-022-00852-0.

[11] Medtronic, Azure™ S SR MRI SureScan™ W3SR01, (2023).

[12] S. Palsaniya, H.B. Nemade, A.K. Dasmahapatra, Size dependent triboelectric nanogenerator and effect of temperature, in: 2018 3rd International Conference on Microwave and Photonics (ICMAP), 2018: pp. 1–2. https://doi.org/10.1109/ICMAP.2018.8354615.
[13] Q. Wang, M. Chen, W. Li, Z. Li, Y. Chen, Y. Zhai, Size effect on the output of a miniaturized triboelectric nanogenerator based on superimposed electrode layers, Nano Energy. 41 (2017) 128–138. https://doi.org/10.1016/j.nanoen.2017.09.030.
[14] X. Chen, Z. Zhang, S. Yu, T.-G. Zsurzsan, Fringing effect analysis of parallel plate capacitors for capacitive power transfer application, in: 2019 IEEE 4th International Future Energy Electronics Conference (IFEEC), IEEE, 2019: pp. 1–5.
[15] J. Shao, Y. Yang, O. Yang, J. Wang, M. Willatzen, Z.L. Wang, Designing Rules and Optimization of Triboelectric Nanogenerator Arrays, Advanced Energy Materials. 11 (2021) 2100065. https://doi.org/10.1002/aenm.202100065.
[16] Z. Zhao, L. Zhou, S. Li, D. Liu, Y. Li, Y. Gao, Y. Liu, Y. Dai, J. Wang, Z.L. Wang, Selection rules of triboelectric materials for direct-current triboelectric nanogenerator, Nat Commun. 12 (2021) 4686. https://doi.org/10.1038/s41467-021-25046-z.
[17] N. Wang, J. van Turnhout, R. Daniels, C. Wu, J. Huo, R. Gerhard, G. Sotzing, Y. Cao, Ion-Boosting the Charge Density and Piezoelectric Response of Ferroelectrets to Significantly High Levels, ACS Appl. Mater. Interfaces. 14 (2022) 42705–42712. https://doi.org/10.1021/acsami.2c12185.
[18] K. a. Engebretsen, J. d. Johansen, S. Kezic, A. Linneberg, J. p. Thyssen, The effect of environmental humidity and temperature on skin barrier function and dermatitis, Journal of the European Academy of Dermatology and Venereology. 30 (2016) 223–249. https://doi.org/10.1111/jdv.13301.
[19] V. Nguyen, R. Yang, Effect of humidity and pressure on the triboelectric nanogenerator, Nano Energy. 2 (2013) 604–608. https://doi.org/10.1016/j.nanoen.2013.07.012.
[20] A.V.R. Telang, The influence of rain on the atmospheric-electric field, Terrestrial Magnetism and Atmospheric Electricity. 35 (1930) 125–131. https://doi.org/10.1029/TE035i003p00125.
[21] R. Gunn, The electricity of rain and thunderstorms, Terrestrial Magnetism and Atmospheric Electricity. 40 (1935) 79–106. https://doi.org/10.1029/TE040i001p00079.
[22] A. Karagioras, K. Kourtidis, A Study of the Effects of Rain, Snow and Hail on the Atmospheric Electric Field near Ground, Atmosphere. 12 (2021) 996. https://doi.org/10.3390/atmos12080996.
[23] S. Niu, Y. Liu, S. Wang, L. Lin, Y.S. Zhou, Y. Hu, Z.L. Wang, Theoretical Investigation and Structural Optimization of Single-Electrode Triboelectric Nanogenerators, Advanced Functional Materials. 24 (2014) 3332–3340. https://doi.org/10.1002/adfm.201303799.
[24] K.P.P. Pillai, Fringing field of finite parallel-plate capacitors, Proceedings of the Institution of Electrical Engineers. 117 (1970) 1201–1204. https://doi.org/10.1049/piee.1970.0232.
[25] J. Shao, M. Willatzen, Z.L. Wang, Theoretical modeling of triboelectric nanogenerators (TENGs), Journal of Applied Physics. 128 (2020) 111101. https://doi.org/10.1063/5.0020961.